\documentclass[a4paper]{spie}  
\usepackage{graphicx}
\usepackage{subcaption}
\usepackage[final]{pdfpages}
\usepackage{multirow}
\usepackage{amsmath,amsfonts,amssymb}
\usepackage[colorlinks=true, allcolors=blue]{hyperref}

\title{Commissioning and on-sky performance of FiberPol: a fiber-fed spectropolarimetric system for the SAAO 1.9~m telescope}

\author[a,b]{Siddharth Maharana}
\author[a,c]{K. C. J. Dennison-Farrar}
\author[a,d,e]{Sabyasachi Chattopadhyay}
\author[a]{Nikita Rawat}
\author[a,c]{Nidhi Mehandiratta}
\author[a]{J.C. Viljoen}
\author[f,g]{Matthew Bershady}
\author[a]{David Buckley}
\author[a]{Lisa A. Crause}
\author[a]{Hitesh Gajjar}
\author[a]{Stephen Potter}
\affil[a]{South African Astronomical Observatory, 1 Observatory Rd., 7935, Cape Town, South Africa}
\affil[b]{Sub-department of Astrophysics, University of Oxford, Denys Wilkinson Building, Keble Road, Oxford OX1 3RH, UK}
\affil[c]{Department of Astronomy, University of Cape Town, Private Bag X3, Rondebosch 7701, South Africa}
\affil[d]{Space Science and Engineering Initiative, College of Engineering, University of Hawaii at Manoa, 2540 Dole Street, Honolulu, HI 96822, USA}
\affil[e]{Centre for Space Research, North-West University, Potchefstroom 2520, South Africa}
\affil[f]{University of Wisconsin–Madison, 475 N. Charter St., Madison, WI, USA}
\affil[g]{National Science Foundation, 401 Dulany Street, Alexandria, VA 22314, USA}

\authorinfo{Further author information: (Send correspondence to S.M. and K.D.)\\S.M.: E-mail: siddharth@saao.ac.za, sidh345@gmail.com\\K.D.: cdennisonfarrar@gmail.com}
\begin{document} 
\maketitle

\begin{abstract}
\par At the South African Astronomical Observatory (SAAO), we have developed FiberPol, a fiber-fed spectropolarimetric front-end for the SpUpNIC spectrograph on the 1.9~m telescope. In conjunction with SpUpNIC, it combines two niche techniques: fiber-based spectroscopy and polarimetry to enable a wide range of science cases, particularly for studies of the interstellar medium, and serves as a low-cost technology pathfinder for larger facilities such as the 10~m Southern African Large Telescope.

\par Commissioned in early 2025, FiberPol has already demonstrated key performance benchmarks. Based on an initial data reduction pipeline, we find that the uncertainties in the measured $q$ and $u$ Stokes parameters---and hence the degree of polarization $p$---are typically in the range of 0.2\%--0.3\% in 5~nm spectral bins across the 400--700~nm range. With improved calibration and refined analysis, we expect to reach our design goal of 0.1\% accuracy per bin. Several science programs are already underway following its successful commissioning.

\par FiberPol employs a Wollaston prism and half-wave plate system to analyze linear polarization and extract spectropolarimetric data comprising wavelength ($\lambda$) and the Stokes parameters ($I$, $q$, $u$). Its compact and modular design uses primarily small, commercially available off-the-shelf optical and optomechanical components, making it straightforward to adapt to other spectrographs, including those on larger and next-generation telescopes.

\par As high-accuracy and broadband fiber-fed polarimetry remains relatively unexplored in astronomy, we conducted extensive laboratory tests prior to telescope deployment. These tests confirmed the feasibility of achieving a polarimetric accuracy of 0.1\% for point sources across the 400--700~nm range.

Together, the lab validation and on-sky results demonstrate the viability of high-accuracy fiber-fed spectropolarimetry, opening the path for integrating polarimetric capabilities into conventional spectrographs, and making spectropolarimetry accessible to a broader astronomical community. This paper presents the design, laboratory validation, and commissioning results of FiberPol, and outlines planned upgrades to further enhance its scientific impact.
\end{abstract}

\keywords{polarization, polarimetric modeling, polarimetric calibration, linear polarimetry, optical polarization, fibers, fiber-polarimetry}

\section{Introduction}
Polarimetry refers to the quantitative measurement of the presence and amount of a preferred polarization state in the light reaching us from astronomical sources. While spectroscopy and imaging probe the first and second of the three fundamental properties of light---intensity, frequency, and polarization---polarimetry probes the third one, often providing information that cannot be obtained from imaging and spectroscopy. It is employed by astronomers to investigate the physical properties of myriad astrophysical systems, including active galactic nuclei (AGNs), supernovae, protoplanetary disks, and interstellar dust\cite{Hough_review, trippe2014polarization, Scarrott-1991, andersson_review}. Polarimetry is particularly valuable for studying objects that have a systematic asymmetry in their emission, absorption, or scattering processes. In many such systems, including supernovae, and magnetic fields in the interstellar medium (ISM)\cite{agn_unification, supernova_polarimetry_review, andersson_review}, it provides the only means of inferring the geometry of a system, as often such information is not accessible through imaging or spectroscopic techniques. 

Over the years, astronomers have developed polarimeters with steadily improving accuracies, with state-of-the-art instruments now achieving polarimetric accuracies of 0.1\% or better in the optical regime \cite{robopol, Bailey_2020, DIPOL2}. Most modern optical polarimeters fall into two broad categories: imaging polarimeters, which produce maps of the polarization state across extended fields in broadband filters, and spectropolarimeters, which provide polarization measurements as a function of wavelength. Analogous to the superior diagnostic power of spectra over photometry, spectropolarimetry enables deeper physical insights, and has played a critical role in addressing complex astrophysical questions, such as understanding relativistic jets and the unified model of AGNs \cite{agn_unification}, to the nature of dust in the ISM and its interactions with magnetic fields \cite{andersson_review}.

Most optical polarimeters available today are imagers; very few offer spectropolarimetric capabilities. Although there are many pressing questions in astronomy, especially concerning the physics of dust and magnetic fields\cite{andersson_review} in the ISM that require spectropolarimetry to be addressed, it remains a niche technique with limited availability due to the high cost and complexity of building such systems, as well as challenges in calibrating them to the high accuracies needed for most science cases. Conventional polarimeters are typically specialized and expensive instruments that are challenging to design, build, and maintain because of their large, fragile optics and complex optical designs. Most astronomical sources emit only a few percent or less of polarized light. Achieving meaningful measurements often requires accuracies of $\sigma_p \leq 0.3\%$ per spectral bin to reliably recover the wavelength dependent polarization curve. Consequently, many small telescopes, and almost all large observatories, lack polarimetric capabilities. 

To address the scientific need for high-accuracy spectropolarimetric systems and the technical challenge of building accessible polarimetric systems, at South African Astronomical Observatory's (SAAO) FiberLab, we have built FiberPol. FiberPol is a 
polarimetry-capable fiber-fed front-end for the existing SpUpNIC \cite{SpUpNIC} spectrograph on the SAAO’s 1.9 m telescope. SpUpNIC is a general-purpose, 2 arc-minute long-slit spectrograph with a grating
suite covering the 350–1000 nm wavelength range and spectral resolutions between 500 and 6000. FiberPol is a compact ($20\times20\times20~cm^{3}$), low-cost (under \$10,000), and modular system for the SpUpNIC spectrograph, and is easily mountable or removable from SpUpNIC as needed. FiberPol enables polarimetry by using small, cost-effective fiber-fed units to preprocess polarization signals without requiring any changes to the telescope or downstream spectrograph.

FiberPol was successfully commissioned in March 2025, and multiple science programs are currently underway. Overall, we find that the uncertainties in the measured Stokes parameters $q$ and $u$, and consequently in the degree of polarization $p$, are typically in the range of 0.2\%--0.3\% across the 400--700~nm wavelength range when the data are binned to 5~nm.

The main scientific goal of FiberPol is to enable the acquisition of high-fidelity spectra of the Stokes parameters $I$, $q$, and $u$ in the optical wavelengths of stars for studies of interstellar dust in the ISM. While FiberPol already operates as a scientific instrument, it also serves as a proof of concept for incorporating polarimetric capabilities into a broad range of astronomical instruments and scientific programs. Just as optical fibers have transformed conventional spectroscopy, their integration with spectropolarimetric systems has the potential to enable entirely new observational capabilities.

In later sections, we discuss the opportunities opened up by fiber-fed polarimetry, particularly its role in enabling fiber-based integral-field spectropolarimetry. Although the long-term goal is for FiberPol to operate as a true integral-field unit (IFU) spectropolarimeter for extended sources, this paper presents results from its single-object mode, which was validated during commissioning. Notably, to the best of our knowledge, this is the first instance of a conventional long-slit spectrograph being successfully converted into a fiber-fed spectropolarimeter. Prior to this, HARPSpol was implemented by adding a polarimetric module upstream of the already fiber-fed HARPS spectrograph\cite{HARPSPol}. Fiber-fed polarimetry is a key enabling technology for IFU-based imaging spectropolarimetry and, equally importantly, broadens the range of potential applications. For example, it offers a pathway for adding spectropolarimetric capabilities to existing instruments and facilities.

In this paper, we present the design of the instrument and the results from its on-sky commissioning. Section~\ref{instrument_design} introduces the top-level concept and design of FiberPol. Sections~\ref{lab_ass} and \ref{commissioning} describe the laboratory characterization and commissioning results, respectively. Section~\ref{data_reduction} outlines the current data-reduction pipeline and initial results. Finally, in Section~\ref{concusion}, we discuss future upgrade pathways and the broader scientific impact of fiber-fed spectropolarimetry.

\section{FiberPol Instrument Design}\label{instrument_design}

\begin{figure}
    \centering
    \includegraphics[width=0.67\linewidth]{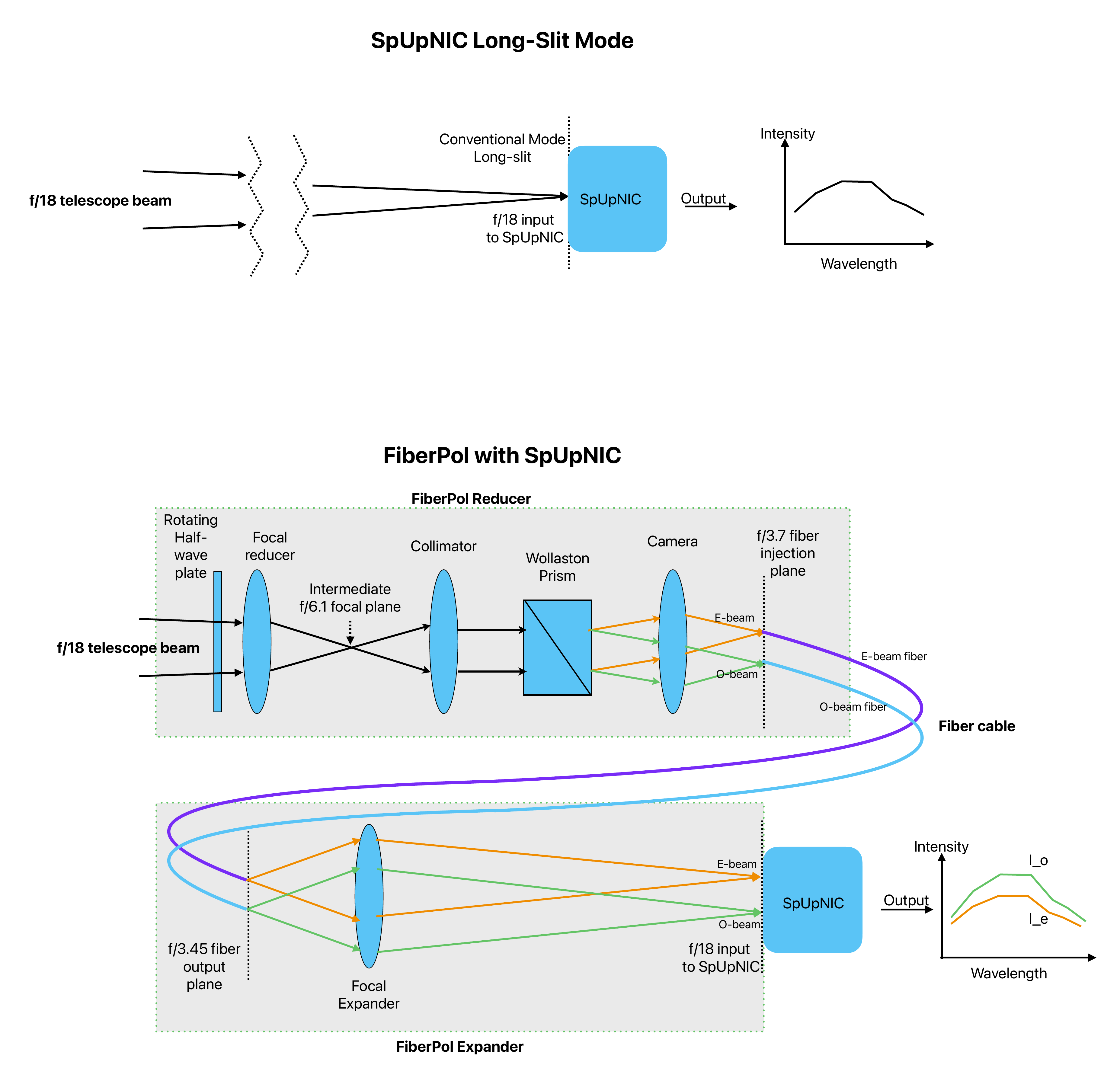}
    \caption{Top level working of FiberPol system when interfaced with SpUpNIC system. }
    \label{fiberpol_schematic}
\end{figure}

The instrument design for FiberPol has been presented in detail in a 2024 SPIE proceedings article\cite{FiberPol_SPIE_2024}, referred to as Paper I from hereon, covering both its optical and optomechanical designs. Aside from a few minor lens additions to aid assembly and mounting on the telescope, the final implementation closely follows the originally proposed design. A top-level schematic of FiberPol and its working principle is shown in Figure~\ref{fiberpol_schematic}. In Paper I, we present a discussion on the rationale behind this system architecture and comparison with other candidate system architectures, keeping into consideration that the eventual goal of FiberPol is to function as an integral-field spectropolarimeter.

\begin{figure}
\begin{subfigure}{0.38\textwidth}
    \centering
    \includegraphics[width=1\linewidth]{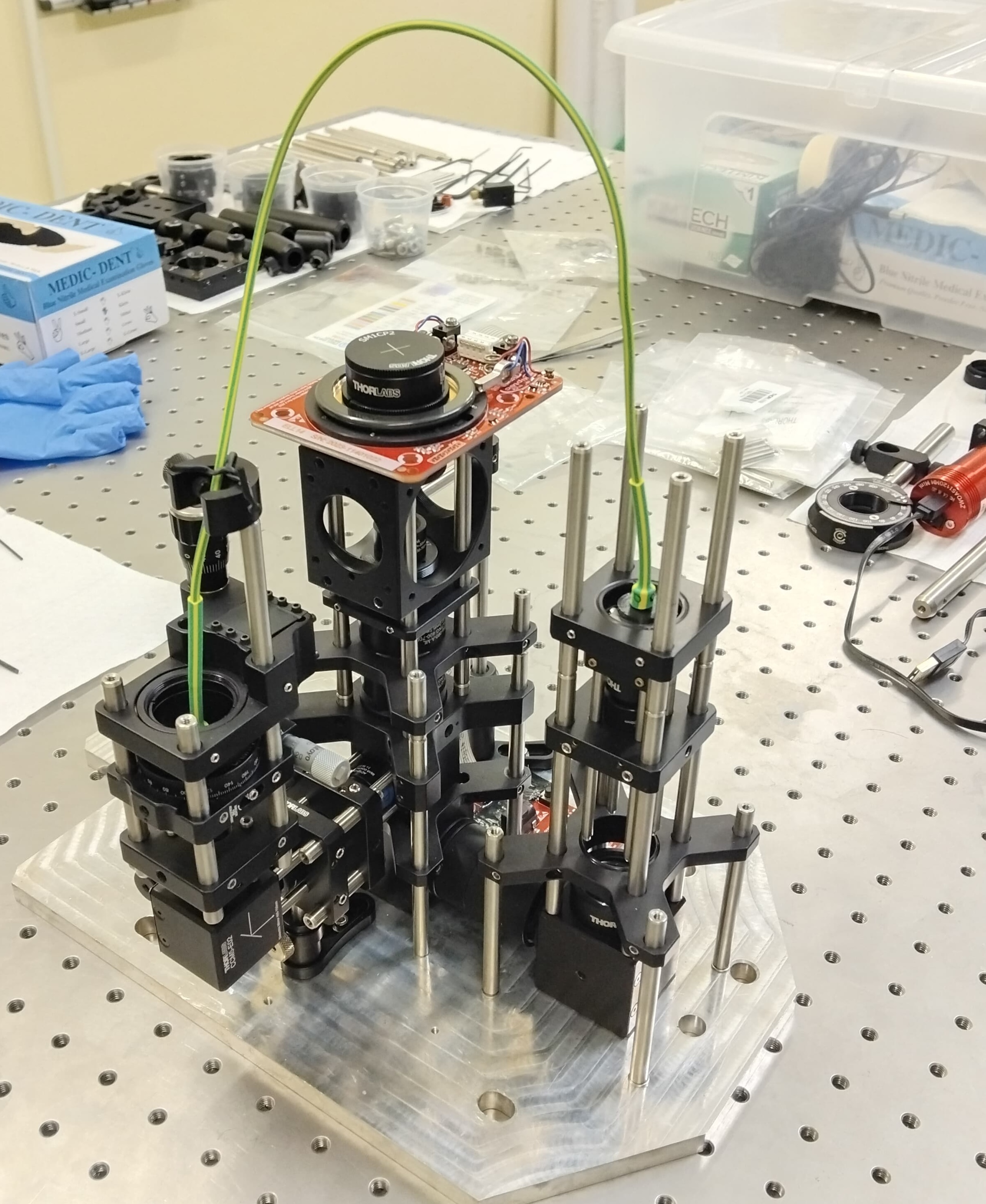}
    \caption{}
    \label{FiberPol_lab_pic}
\end{subfigure}
\begin{subfigure}{0.615\textwidth}
    \centering
    \includegraphics[width=1\linewidth, angle=180]{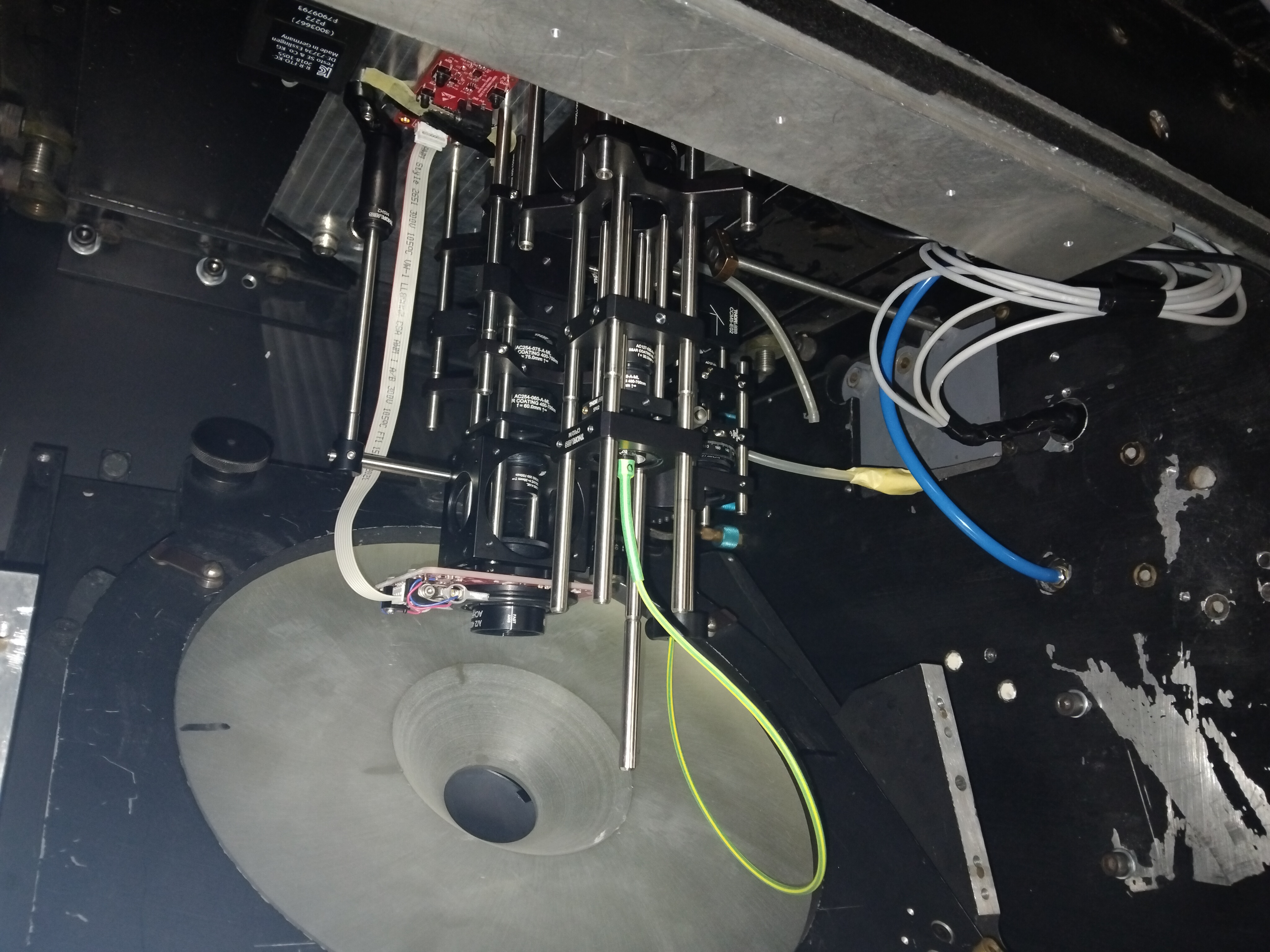}    
    \caption{}
    \label{FiberPol_ontelescope_pic}
\end{subfigure}
\caption{Left: FiberPol fully aligned and assembled in the lab before commissioning. Right: FiberPol mounted on the telescope inside the acquisition box of SpUpNIC. As can be noticed, the 45 deg fold mirror imposes very tight spatial constraints, which FiberPol is able to conform to. }
\end{figure}

FiberPol is installed in the limited volume available inside the SpUpNIC acquisition box, located optically between the telescope and SpUpNIC (Figure~\ref{FiberPol_ontelescope_pic}). In its usual long-slit mode, SpUpNIC is fed an f/18 beam at the telescope's Cassegrain focus. The details of the telescope and SpUpNIC are listed in Table~\ref{SpUpNIC_spec}. FiberPol has been designed to introduce fiber-fed spectropolarimetry while preserving the f/18 input beam to SpUpNIC. The optical system enables two-channel linear spectropolarimetry using a rotating half-wave plate (HWP) and a Wollaston prism (WP) as the polarization analyzer. The two orthogonal polarization beams, commonly referred to as the ordinary ($o$) and extraordinary ($e$) beams, are injected into dedicated fibers, along with an additional sky-fiber for measuring the sky background. These fibers are reformatted into a pseudo-slit that feeds into SpUpNIC, maintaining compatibility with the spectrograph’s usual long-slit format input. As a result, each exposure produces three spectra: one each from the $o$, $e$, and sky fibers (as shown in Figure~\ref{on-sky_exposure}). The linear polarization state is derived by computing the normalized difference between the sky subtracted $o$ and $e$ spectra across multiple HWP angles, yielding wavelength-dependent Stokes parameters. In addition to this key task of polarization dependent beam splitting, the FiberPol system carries out other important tasks, which are described later.

\begin{table}[!htbp]
\centering
\caption{FiberPol host: details of the SAAO 1.9 m telescope and SpUpNIC spectrograph}.
\begin{tabular}{ccc}
\hline
\textit{Sl. No.} & \textit{Parameter}                & \textit{Value}                                   \\ \hline
1 & Telescope f/\#           & 18.0                                    \\
2 & Telescope Aperture       & 1.88 m                                  \\
3 & Slit length              & 2’ ($\sim$37 mm)                        \\
4 & Median seeing FWHM  & 1.5"\\
5 & Telescope Port           & Cassegrain                 \\
6 & SpUpNIC Wavelength Range & 350-1000 nm                       \\
7 & Spectral Resolution      & 500 to 6500                             \\
8 & No. of gratings          & 10                                      \\ \hline
\end{tabular}
\label{SpUpNIC_spec}
\end{table}

\begin{table}[!htbp]
    \centering
    \caption{FiberPol system design goals}    \begin{tabular}{ccc}
        \hline
        \textit{Parameter} & \textit{Technical Goal} \\
        \hline
        Polarimetric Accuracy  & 0.1~\%\\
        Polarimeter Type & Two Channel Linear Polarimetry \\
        Analyzer Type & WP+HWP \\
        Field of View & $10"\times20"$\\
       PSF Performance & Seeing limited at fiber-injection plane\\
        f/\# at fiber feed  & 3.5 to 4.5 \\
        Telecentricity & $\leq0.1^{\circ}$ \\
        Wavelength Range & 400~nm -- 700~nm\\
        Components & Off-the-shelf \\
        Cost & Under $10,000\$$\\
        Fiber Aperture & $100/140~{\mu}m$ core/cladding \\
        \hline
    \end{tabular}
    \label{tech_goals}
\end{table}

The instrument consists of three major optical subsystems: the \textit{Reducer}, the \textit{Fiber Cable} and the \textit{Expander}. The 1.9~m Telescope delivers a slow $f/18$ beam to SpUpNIC/FiberPol entrance, which is not optimal for fiber injection due to significant focal-ratio degradation (FRD) for such slow beams. The \textit{Reducer} acts as a focal reducer, converting the telescope’s native $f/18$ beam to $f/3.75$ for efficient fiber injection and minimizing FRD. More importantly, beam splitting into the ordinary ($o$) and extraordinary ($e$) polarization channels occurs within the \textit{Reducer}. A collimated beam is directed through a WP, which performs the polarization separation. Polarimetric modulation---used to measure the linear Stokes parameters and mitigate instrumental polarization---is introduced via a rotating HWP mounted on a motorized rotary stage at the very entrance of the instrument.

\begin{figure}
\begin{subfigure}{0.5\textwidth}
    \centering
    \includegraphics[width=1.0\linewidth]{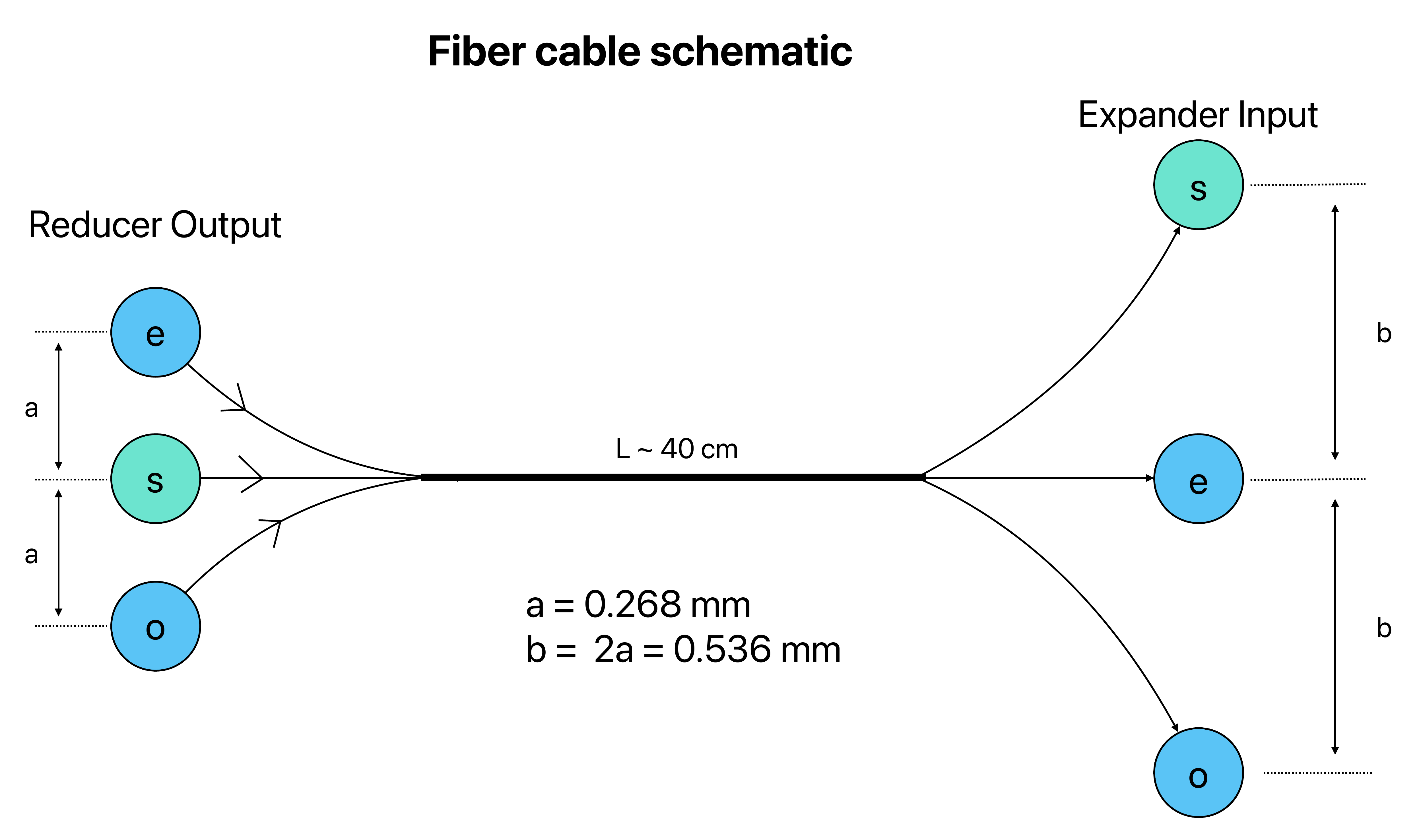}
    \caption{} 
    \label{fiber_layout}
\end{subfigure}
\begin{subfigure}{0.5\textwidth}
    \centering
    \includegraphics[width=0.9\linewidth]{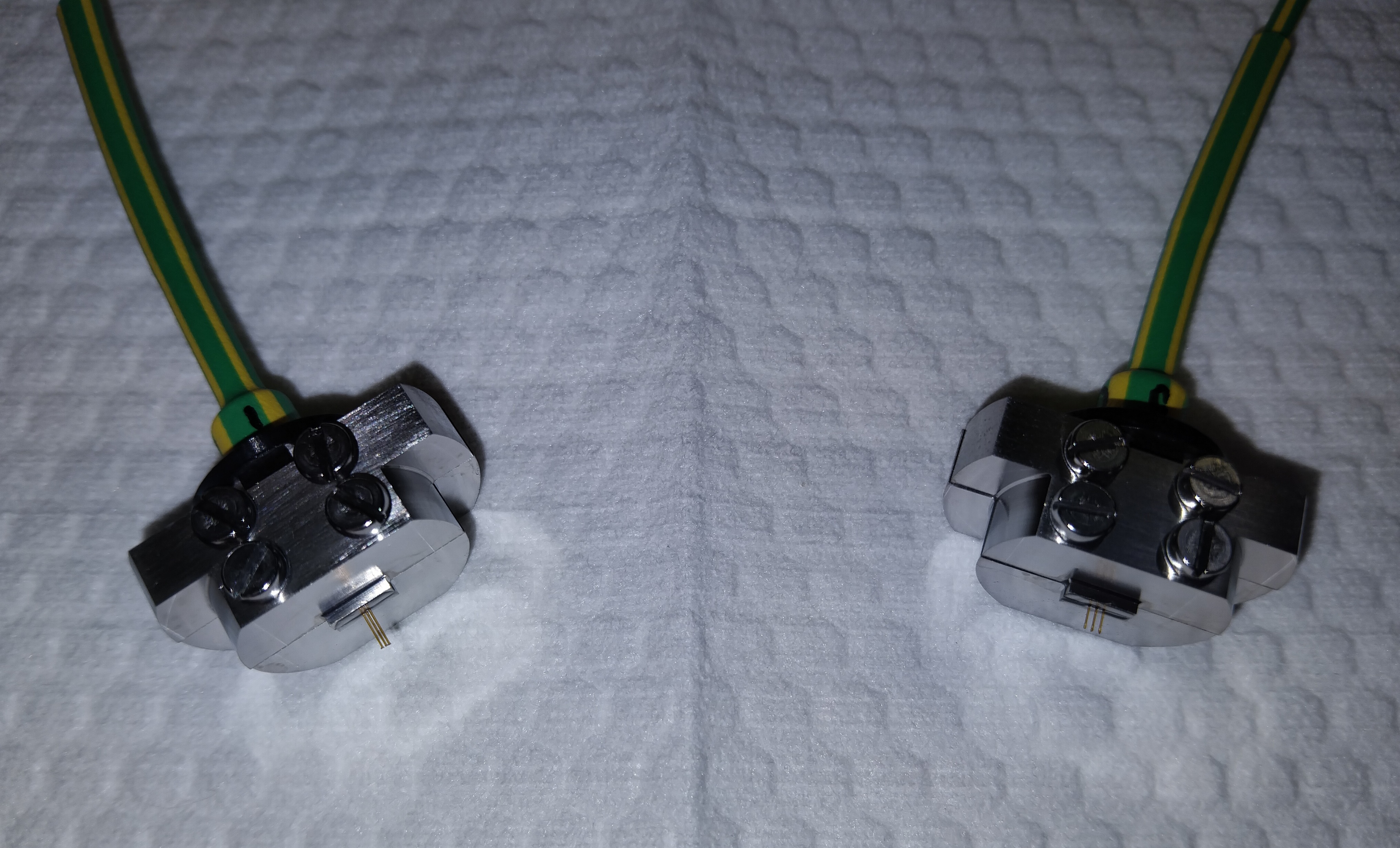}
    \caption{}
    \label{FiberPol_fiber_ass}
\end{subfigure}
\caption{Left: Schematic showing the fiber arrangement in the custom-built cable for FiberPol. At the input end, the ordinary ($o$) and extraordinary ($e$) fibers are separated by 0.536~mm, with a sky fiber placed in between to enable simultaneous sky background measurement. At the output end, the $e$, $o$, and sky fibers are spaced further apart to avoid cross-talk between their spectral traces. Right: Photograph of the fabricated fiber cable (prior to polishing), showing fiber terminations at both ends.}
\label{fiber_cable}
\end{figure}

The expected fiber output for an $f/3.75$ input beam is approximately $f/3.45$\cite{FRD_Sabyasachi_paper}. This beam is subsequently expanded back to $f/18$ by the \textit{Expander} to match the input requirements of SpUpNIC. The \textit{Fiber Cable}, whose schematic is shown in Figure~\ref{fiber_cable}, routes the beam from the \textit{Reducer} to the \textit{Expander}.

\subsection{Technical Requirements}

The top-level technical requirements that guided the design of FiberPol were:

\begin{enumerate}
\item No modifications to the SpUpNIC spectrograph.
\item No modifications to the telescope hardware.
\item No change to the optical input beam of SpUpNIC.
\item Mitigation of the polarization-scrambling behavior of multi-mode fibers.
\item Fit within a volume of roughly $20\times 20 \times 20~cm^{3}$ available inside the SpUpNIC acquisition box.
\end{enumerate}

FiberPol meets all the above requirements and operates as a front-end system that is fully compatible with the existing telescope and spectrograph infrastructure. It is a compact, cost-effective system built entirely from commercial off-the-shelf optical and optomechanical components—with the sole exception of a custom fiber ferrule, which was fabricated in the SAAO workshop. This enabled the rapid development and deployment of the instrument. Achieving the required performance with such components necessitated judicious design and analysis, and careful laboratory alignment and verification. The complete optical prescription, including the specific components used in the FiberPol system, is presented in Paper I.

The optical system was optimized for the wavelength range of 400--700 nm for the following two reasons: (a) this range is well matched to the availability of commercially available optical components with high-performance anti-reflection coatings, and (b) it is centered around the typical peak of the interstellar polarization curve of 550 nm, described by the Serkowski function\cite{andersson_review}.

Importantly, in FiberPol, light is injected into the fibers only after the polarization analysis has been performed by the WP. This design choice ensures that the inherent polarization-scrambling behavior of multi-mode fibers is mitigated. From this point onward, the primary polarization effect introduced by the fibers and downstream optics is differential throughput between the two orthogonal polarization beams, like any other conventional optical element. This can be corrected through careful calibrations, which are discussed in Section~\ref{data_reduction}.

\section{Assembly and Characterization in the lab}\label{lab_ass}

Figure~\ref{FiberPol_lab_pic} shows the fully assembled FiberPol system in the lab. The assembly was carried out in multiple phases to ensure a modular and phased integration of the subsystems and precise optical alignment. On the telescope, the \textit{Reducer} subsystem of FiberPol is fed by the telescope’s native $f/18$ beam. To replicate this in the lab, a custom f/18 beam was created to emulate the expected input beam. 

\par The first test conducted was to verify the separation of the $o$ and $e$ beams at the reducer/fiber-injection focal plane. The optical design predicts a nominal separation of 0.536~mm, and the measured separation was found to be $0.536 \pm 0.008$~mm, confirming that the assembled system performs according to design expectations.

Following this, the \textit{Expander} and \textit{Fiber Cable} subsystems were integrated. At each step, careful alignment and verification tests were performed. A key challenge was achieving the required $536~\mu$m separation between the $e$ and $o$ fibers with $<5~\mu$m precision. This demanded custom fabrication of a precise fiber ferrule, which was accomplished iteratively in close collaboration with the SAAO mechanical workshop. Fiber mounting, gluing, and polishing were carried out to achieve an optical-quality finish. Fiber FRD and throughput measurements were also performed to ensure that the fiber system met performance requirements. We do find that the two science and the sky fibers have a differential throughput between them, but this can be easily corrected by post-processing techniques, as has been detailed in Section~\ref{data_reduction}.

Another critical step was aligning the \textit{Fiber Cable} with the optical system of FiberPol to within $5~\mu$m accuracy, especially matching the \textit{Fiber Cable} axis to the WP axis. This was accomplished using a dedicated lab setup and careful iteration. After successful integration, FiberPol underwent extensive lab testing to characterize instrumental systematics and assess its polarimetric performance in fiber-fed mode. The system demonstrated polarimetric accuracy better than 0.1\% across the 400–700~nm wavelength range.

\section{Telescope Commissioning}\label{commissioning}
\subsection{Integration with SpUpNIC and the 1.9~m telescope}
As noted, {FiberPol was designed to interface with the existing SpUpNIC spectrograph without requiring any modifications to the instrument itself—except for the removal of the plate that holds SpUpNIC’s long-slit mechanism. In its place, an exact mechanical replica (the FiberPol mounting plate) was installed. The USB-cable for controlling the HWP rotation mechanism in FiberPol was connected via a telescope-side port connected to the SpUpNIC PC in the \textit{telescope warm-room} and controlled using a Python script. All other operations and observations with FiberPol were performed using the standard SpUpNIC control software.

To align FiberPol with the instrument and telescope, the following procedure was followed:

\begin{enumerate}
    \item \textit{Installation and Wavelength Calibration:} 
    FiberPol was installed onto SpUpNIC by replacing the original slit plate with the FiberPol plate. Prior to the installation, arc lamp exposures were taken using the grating settings intended for use with FiberPol (Grating 6 that allows 420--680~nm range). Since the current FiberPol setup does not support arc lamp illumination, these pre-recorded arcs were used to establish the initial/first-guess wavelength calibration for science exposures.

    \item \textit{Alignment Using Rear-of-Slit Mirror (ROS) camera:}
    SpUpNIC includes a rear-of-slit mirror (ROS) camera, which, in the absence of the slit, focuses on the nominal focal plane where the re-imaged $f/18$ beam from the fibers is expected to fall. This allowed us to verify that the fiber pseudo-slit was properly aligned and parallel to the original slit axis. An image of the aligned fiber pseudo-slit as seen through the ROS camera is shown in Figure~\ref{flat_exposure}.

    \item \textit{Flat Field and Focus Check:} 
    Dome flat exposures were taken to confirm (a) the relative position of the fiber traces on the detector, and (b) to determine the optimal camera focus with FiberPol installed.
\end{enumerate}

All the above tasks were completed during the daytime. As a testament to the robustness of FiberPol’s design, no adjustments to either the telescope or camera focus were required during the installation, confirming the accuracy of the optical design and alignment. Furthermore, science targets were successfully observed on the very first night of commissioning, demonstrating that FiberPol integrated seamlessly with both SpUpNIC and the telescope system.

\begin{figure}
\begin{subfigure}{0.99\textwidth}
    \centering
    \includegraphics[width=1.0\linewidth]{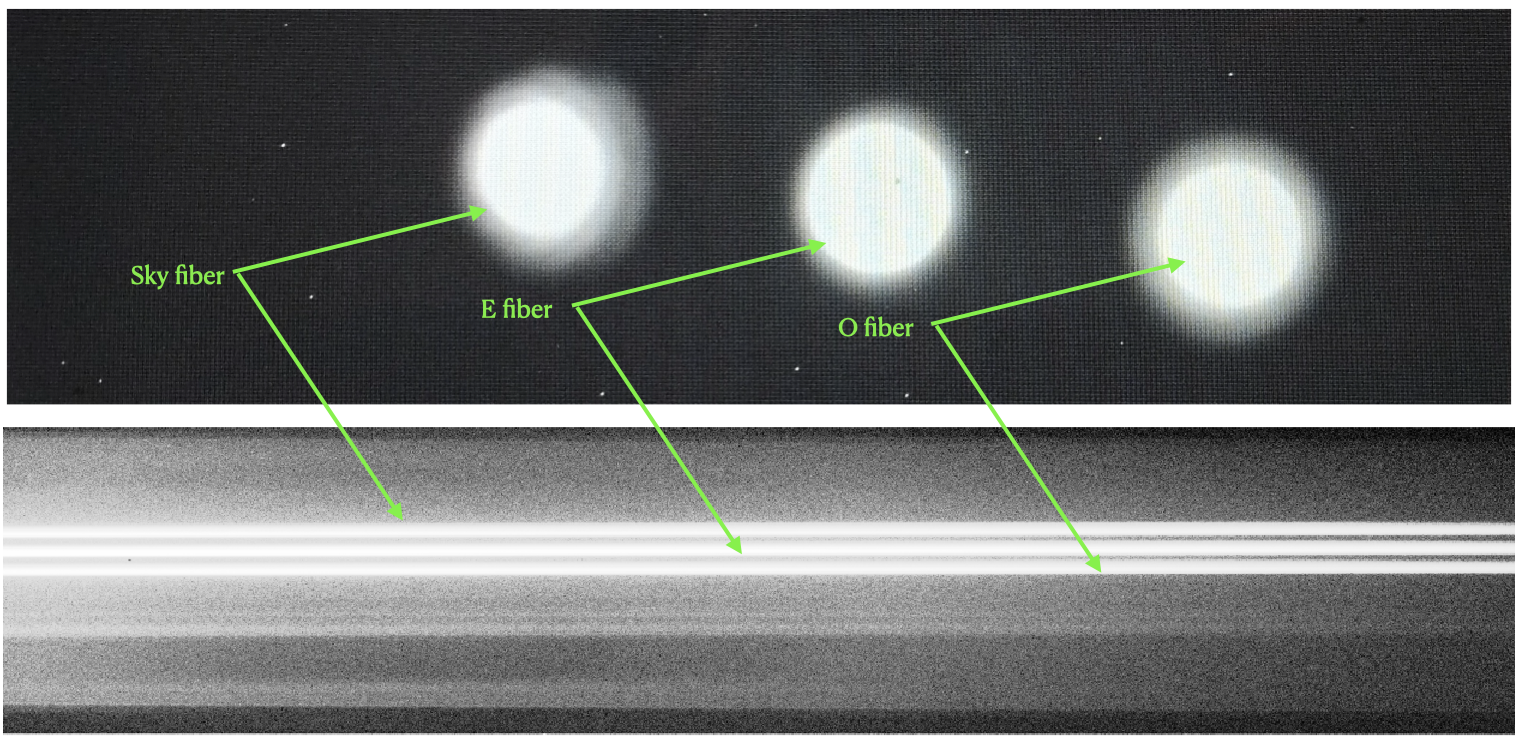}
    \caption{}
    \label{flat_exposure}
\end{subfigure}

\begin{subfigure}{0.99\textwidth}
    \centering
    \includegraphics[width=1.0\linewidth]{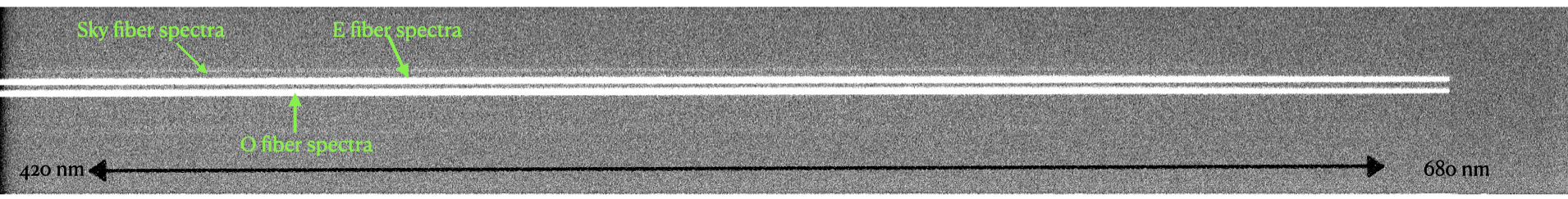}
    \caption{}
    \label{science_exposure}
\end{subfigure}
\caption{Spectra from FiberPol. Top panel: Flat field image of the fibers as seen from the ROS camera. Middle panel: Flat field exposure showing the three fiber traces corresponding to the $e$, $o$, and sky fibers. Bottom panel: Example science exposure of a star showing the corresponding $e$ and $o$ spectra, along with the sky background. The intensity differences between the $e$ and $o$ spectra across HWP positions are used to derive the Stokes parameters.}
\label{on-sky_exposure}
\end{figure}

\subsection{Night-time Observations}

Once the daytime alignments and integration of FiberPol with SpUpNIC were completed successfully, the instrument was ready for science observations. The first step during on-sky operation was to identify the fiber position on the sky using the SpUpNIC acquisition camera. This process involves determining the so-called \textit{magic pixel}—the pixel location on the acquisition camera corresponding to the star's position that aligns with the input fibers. This location is then used to acquire subsequent science targets efficiently.  Figure~\ref{science_exposure} shows a typical FiberPol science image, consisting of the $e$, $o$ and the sky spectra. 

For conducting observations with FiberPol, the standard SpUpNIC control software and GUI are used in conjunction with a Python script that currently controls the HWP rotation mechanism separately. Complete integration of HWP control into the SpUpNIC software is planned for the future upgrades.

Target acquisition is done by first aligning a bright star (typically a polarimetric standard) onto the magic pixel. Over time, minor shifts in the true magic pixel location may occur due to flexure or alignment changes. Nevertheless, it provides a reliable starting point for fine-tuning the target position using the telescope’s hand-paddle/fine-guiding controls. Once the star is accurately placed on the fibers, auto-guiding is initiated, and science exposures commence. This semi-manual method of target acquisition is prone to errors, and is likely a dominant contributor to the current uncertainty limit of the instrument. 

A typical observation sequence for a target consists of 3–5 cycles of exposures at HWP positions of $0^{\circ}$, $22.5^{\circ}$, $45^{\circ}$, and $67.5^{\circ}$. These positions allow for the extraction of multiple independent measurements of the Stokes parameters $q$ and $u$, enabling both internal consistency checks and robust uncertainty estimation.

As part of the commissioning and characterization phase, we observed multiple polarized and unpolarized standard stars, primarily from widely used community catalogs\cite{cikota_standards, southern_standards_gil-hutton}. In addition to these calibration observations, a significant portion of the observing time was dedicated to science targets. These observations aim to measure wavelength-dependent stellar polarization, from which Serkowski parameters can be derived to study dust grain size distributions and magnetic field geometries in various astrophysical environments. 

\section{Data Reduction and Initial Results}\label{data_reduction}

As shown in Figure~\ref{science_exposure}, each exposure yields the $e$ and $o$ spectra from the science target, along with a corresponding sky spectrum. A wavelength solution is applied to the spectra by identifying and correcting for any shifts relative to the initial-guess wavelength calibration (obtained from pre-FiberPol installation arcs) for each object, using spectral features present in the target itself. We find that, within a given observation sequence spanning multiple HWP angles, the wavelength calibration remains stable to within a few~\AA, which is sufficient for our scientific objectives.

The Stokes parameters are derived from the normalized differences in intensities between the $e$ and $o$ spectra across the HWP rotation positions. Specifically, the intensities $I_e$ and $I_o$ were extracted for each HWP angle. To account for the differential transmission between $I_e$ and $I_o$ in the instrument optics and fibers, a correction factor $k$ is applied, as defined in Equations~\ref{kq_factor} and~\ref{ku_factor} for the $q$ and $u$ measurements, respectively. Using this factor, the corrected $o$-beam intensity is computed as $I_o' = \frac{I_o}{k}$.

\begin{equation}\label{kq_factor}
    k_q = \left[\frac{{I_{o}}^{0}\times{I_{o}}^{45}}{{I_{e}}^{0}\times{I_{e}}^{45}}\right] ^\frac{1}{2}
\end{equation}

\begin{equation}\label{ku_factor}
    k_u = \left[\frac{{I_{o}}^{22.5}\times{I_{o}}^{67.5}}{{I_{e}}^{22.5}\times{I_{e}}^{67.5}} \right]^\frac{1}{2}
\end{equation}

The normalized Stokes parameter for a given HWP position is then computed as the normalized difference between the two corrected intensities, as described in Equation~\ref{rm}. The value of $r_m$ corresponds to either $+q$, $-q$, $+u$, or $-u$ depending on the HWP position. The final values of $q$ and $u$ are obtained using Equations~\ref{q_NOT} and~\ref{u_NOT}, respectively. The uncertainties in the measurements are estimated from the standard deviation of the corresponding $r_m$ values obtained across multiple HWP cycles.

A dedicated data reduction pipeline based on IRAF has been developed to analyze the raw data obtained from FiberPol. Here, we present results based on this initial version of the pipeline. We anticipate that the instrument performance and analysis accuracy will improve as the pipeline matures.

\begin{equation}\label{rm}
r_m = \frac{{I_{o}^{'}}- {I_{e}}}{{I_{o}^{'}}+ {I_{e}}}
\end{equation}

\begin{equation}\label{q_NOT}
q = \frac{{r_m}^{0} - {r_m}^{45}}{2}
\end{equation}

\begin{equation}\label{u_NOT}
u = \frac{{r_m}^{22.5} - {r_m}^{67.5}}{2}
\end{equation}

\subsection{Initial Results}

\par To begin, we determine the instrumental polarimetric \textit{zero offsets} using observations of unpolarized standard stars. \textit{Overall, we find that the uncertainties in the measured $q$ and $u$ parameters, and hence in degree of polarization, $p$ lie mostly in the range of 0.2\%--0.3\% across the 400--700~nm range, when computed in 5~nm bins.} This is illustrated in Figure~\ref{inst_pol_fiberpol}. The uncertainties increase toward the extreme ends of the wavelength range due to reduced system throughput. The results shown in Figure~\ref{inst_pol_fiberpol} represent data combined over approximately a week. The error bars in these plots reflect the standard deviation of the measurements and provide a direct estimate of the polarimetric precision of the instrument.

A wavelength-dependent instrumental polarization is observed, with absolute values typically in the range of 0.5\%--1\% for both $q$ and $u$. This behavior most likely arises from the telescope optics. As the telescope lies before the HWP, the polarization introduced by it cannot be corrected with HWP modulation. However, since the instrumental polarization is stable over time, it poses no limitation and can be reliably removed. This derived instrumental zero polarization is subtracted from all subsequent science measurements.

The instrumental \textit{polarization efficiency} and the orientation of the instrument's coordinate system relative to the celestial reference frame are determined using polarized standard stars. We observed multiple such standards and used the mean offset in $q$ and $u$ to compute the effective rotation angle of the instrument frame. This rotation correction, together with the zero-polarization offset, is applied to all measurements.

Figures~\ref{HD298383} and \ref{vela} show the calibrated FiberPol measurements of two polarized standards and their comparison to catalog values. As seen, their FiberPol measurements agree very well within the quoted uncertainties to the catalog values. This also confirms that the polarization efficiency of FiberPol for both $q$ and $u$ is close to 1.


\begin{figure}[!htbp]
    \centering
    \includegraphics[scale=1]{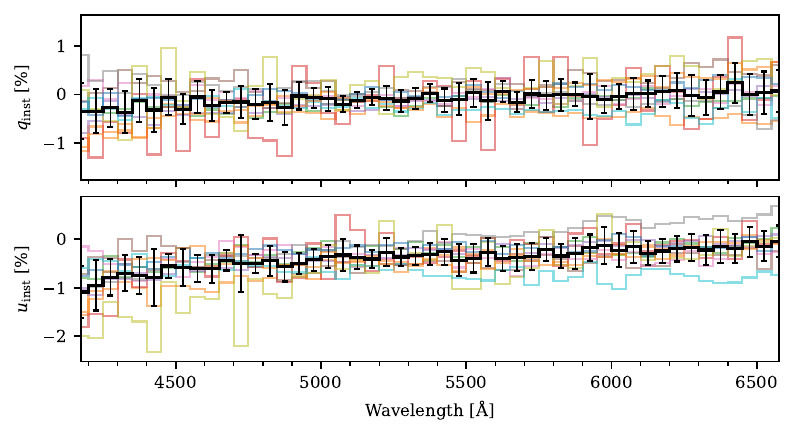}
    \caption{Preliminary instrumental polarization of FiberPol as measured through observations of multiple standard unpolarized stars. As can be noted, the binning is 5~$nm$ bins, and the error-bars shown are the standard deviation of the measurements.}
    \label{inst_pol_fiberpol}
\end{figure}

\begin{figure}[!htbp]
    \centering
    \includegraphics[scale=1]{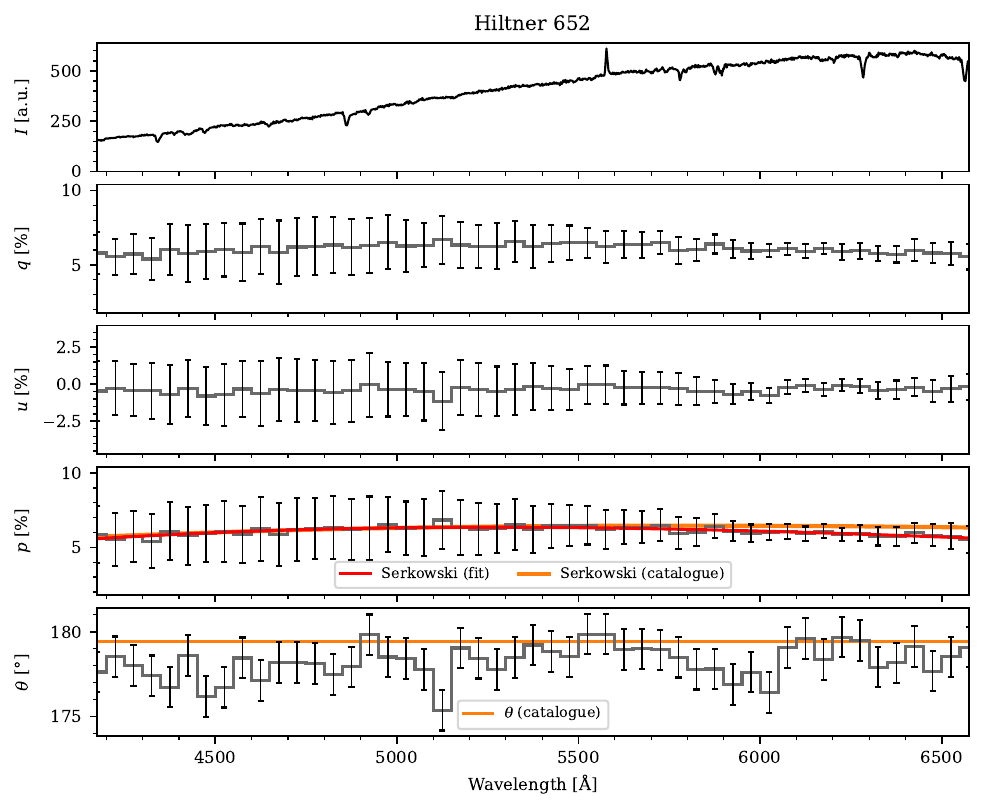}
    \caption{Measured Stokes parameters of a polarized standard star Hiltner 652. The catalog values of $p$ and $\theta$ are overplotted, and thus match very well with our measurements.}
    \label{HD298383}
\end{figure}
\begin{figure}[!htbp]
    \centering
    \includegraphics[scale=1]{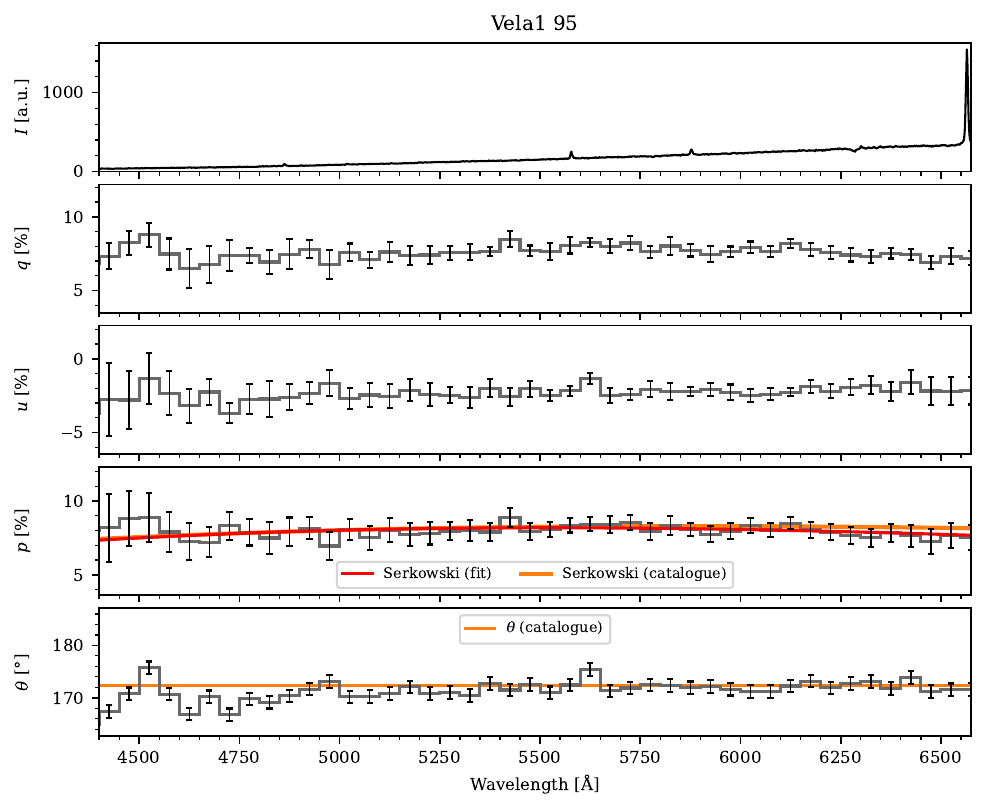}
    \caption{Measured Stokes parameters of a polarized standard star Vela1 95. The catalog values of $p$ and $\theta$ are overplotted, and thus match very well with our measurements.}
    \label{vela}
\end{figure}

\section{Discussion and Future Upgrades}\label{concusion}

Even with a preliminary analysis, we find that FiberPol has already demonstrated the capability to perform high-accuracy polarimetry on-sky. The achieved instrumental accuracy of 0.2\%--0.3\% in 5~nm bins is sufficient for conducting science related to interstellar medium physics. Nevertheless, we aim to refine the analysis further by improving data reduction and post-processing techniques to eliminate any remaining systematic errors. Ongoing efforts are focused on developing a robust and automated pipeline, which will ultimately be released for community use, thereby enabling broader scientific application of FiberPol. The limiting sources of uncertainty for FiberPol likely include a combination of the following:

\begin{enumerate}
    \item \textit{Target acquisition errors:} As described in Section~\ref{commissioning}, the targets are acquired semi-manually on the fibers using the telescope’s hand-paddle controls. Imperfect target acquisition, when combined with seeing and tracking errors described below are likely the predominant source of measurement uncertainty. 
    
    \item \textit{Tracking errors:} Imperfect telescope tracking can cause differential throughput between the $e$ and $o$ beams on the fiber face. This can induce variable instrumental polarization during an observation cycle. 

    \item \textit{Seeing variations:} Changes in atmospheric seeing can broaden or compress the point spread function (PSF) at the fiber injection plane, leading to relative flux changes between the $e$ and $o$ beams on their fibers. 
\end{enumerate}
The last two effects are more pronounced when the $e$ and $o$ beams are not centered on their fibers due to imperfect acquisition. The following upgrades are planned to overcome these limitations and expand FiberPol's capabilities:

\begin{enumerate}

    \item \textit{Fold mirror mechanism for arcs and acquisition:} The most immediate and crucial upgrade we plan is of adding an in-out fold mirror mechanism to FiberPol for the following purposes:
    \begin{enumerate}

        \item To feed light into an internal acquisition camera. This addition would make object acquisition significantly more efficient and reliable, especially for fainter targets. By centering the target's $e$ and $o$ images on the fibers, the effect of seeing and tracking errors would be predominantly nullified.
        
        \item To allow taking arc images after each observation, which is useful for improving wavelength calibration. 

    \end{enumerate}
    
    \item \textit{Integration of HWP control with SpUpNIC software:} This will significantly simplify observing with FiberPol, making it nearly identical to regular SpUpNIC operations. Currently, the HWP is operated via a standalone Python script running on the SpUpNIC PC. 

    \item \textit{Extending wavelength coverage:} Currently, FiberPol is optimized for the 400--700~nm range due to the high-throughput ($>$99\%) anti-reflection coatings on the lenses. However, the system can be extended to a broader range (350--1000~nm) with a modest trade-off in throughput (to $\sim$98\%) of anti-reflection coatings. This extension would require replacing the lenses with ones coated for the full optical range. This extended wavelength range would lead to improved Serkowski function measurement, and thus a significant enhancement in the potential science yield.
\end{enumerate}

\section{Summary and Conclusions}
The commissioning of FiberPol demonstrates that high-accuracy fiber-fed spectropolarimetry can be achieved using a compact, low-cost front-end without modifications to the host telescope or spectrograph. In its initial on-sky tests, \textit{FiberPol} has demonstrated the capability to perform high-accuracy spectropolarimetric measurements. Using a preliminary data reduction pipeline, we achieve a polarimetric precision of 0.2--0.3\% in 5~nm spectral bins across 400--700~nm, as estimated from repeated measurements of different unpolarized standard stars. Measurements of polarized standards agree with catalog values within the quoted uncertainties, demonstrating comparable absolute accuracy. This level of accuracy is consistent with laboratory tests and sufficient for probing interstellar dust and magnetic field structures.

We measured and corrected for \textit{instrumental polarization}, which is found to be wavelength-dependent but stable over time, enabling reliable calibration using standard stars. Observations of \textit{unpolarized standards} established the instrument's baseline \textit{polarimetric zero-offsets}, while polarized standard stars were used to determine the \textit{polarization efficiency} and align the instrument’s reference frame with the celestial coordinate system. The results show broad agreement with catalog values, confirming that FiberPol’s polarization efficiency is close to unity for both $q$ and $u$ parameters.

The current limitations to accuracy likely stem from imperfect target acquisition process combined with seeing variations and telescope tracking errors. Planned upgrades including improved HWP control integration, an internal fold-mirror mechanism for arcs and acquisition, and extension of the wavelength range are expected to further enhance FiberPol’s performance, usability and science yield.

\subsection{Implications for Future Instrumentation}
\begin{itemize}
\item Adaptability: The ability to achieve high-accuracy polarimetry with fiber-fed systems opens up the possibility for retrofitting existing spectrographs, making polarimetry accessible to more observatories.
\item Cost Efficiency: FiberPol’s design, based on commercial off-the-shelf components, provides a low-cost alternative to large, dedicated polarimeters.
\item 2D Spectropolarimetry: The successful demonstration of single-object polarimetry lays the groundwork for developing multi-object and integral field polarimetry, enabling the study of extended astrophysical objects such as galaxies and nebulae.
\item Broad Scientific Applications: This development opens up new possibilities for integrating fiber-fed polarimetric capabilities across a wide range of instruments, enhancing fields such as interstellar dust studies and exoplanet characterization, and expanding the scope of polarimetric observations. For example, fibers can feed bench polarimeters to achieve the extreme precision polarimetry (better than $10^{-5}$) required for exoplanet research\cite{berdyugin_exoplanet}, effectively isolating the system from external influences like temperature fluctuations and gravity-induced flexures—key sources of instrumental variability.
\end{itemize}

\acknowledgments
The authors acknowledge the support of the National Research Foundation of South Africa for this project through the Salt Research Chair grant SARChI-114555. SM thanks Prof A.N. Ramaprakash for useful discussions and Prof. John Thorstensen and Dr. Hannah Worters for their help with commissioning related work. We also acknowledge the support of the technical staff of SAAO and the Sutherland Observatory.

\bibliography{article_bib.bib} 

\begin{thebibliography}{10}

\bibitem{Hough_review}
Hough, J., ``{Polarimetry: a powerful diagnostic tool in astronomy},'' {\em Astronomy And Geophysics}~{\bf 47},  3.31--3.35 (06 2006).

\bibitem{trippe2014polarization}
Trippe, S., ``Polarization and polarimetry: A review,'' (2014).

\bibitem{Scarrott-1991}
Scarrott, S., ``Optical polarization studies of astronomical objects,'' {\em Vistas in Astronomy}~{\bf 34},  163 -- 177 (1991).

\bibitem{andersson_review}
Andersson, B.-G., Lazarian, A., and Vaillancourt, J.~E., ``Interstellar dust grain alignment,'' {\em Annual Review of Astronomy and Astrophysics}~{\bf 53}(1),  501--539 (2015).

\bibitem{agn_unification}
{Antonucci}, R.~R.~J. and {Miller}, J.~S., ``{Spectropolarimetry and the nature of NGC 1068.},'' {\em Astrophysical Journal}~{\bf 297},  621--632 (Oct. 1985).

\bibitem{supernova_polarimetry_review}
Wang, L. and Wheeler, J.~C., ``Spectropolarimetry of supernovae,'' {\em Annual Review of Astronomy and Astrophysics}~{\bf 46}(Volume 46, 2008),  433--474 (2008).

\bibitem{robopol}
Ramaprakash, A.~N., Rajarshi, C.~V., Das, H.~K., Khodade, P., Modi, D., Panopoulou, G., Maharana, S., Blinov, D., Angelakis, E., Casadio, C., Fuhrmann, L., Hovatta, T., Kiehlmann, S., King, O.~G., Kylafis, N., Kougentakis, A., Kus, A., Mahabal, A., Marecki, A., Myserlis, I., Paterakis, G., Paleologou, E., Liodakis, I., Papadakis, I., Papamastorakis, I., Pavlidou, V., Pazderski, E., Pearson, T.~J., Readhead, A. C.~S., Reig, P., Słowikowska, A., Tassis, K., and Zensus, J.~A., ``{RoboPol: a four-channel optical imaging polarimeter},'' {\em Monthly Notices of the Royal Astronomical Society}~{\bf 485},  2355--2366 (02 2019).

\bibitem{Bailey_2020}
Bailey, J., Cotton, D.~V., Kedziora-Chudczer, L., De~Horta, A., and Maybour, D., ``Hippi-2: A versatile high-precision polarimeter,'' {\em Publications of the Astronomical Society of Australia}~{\bf 37} (2020).

\bibitem{DIPOL2}
Piirola, V., Berdyugin, A., and Berdyugina, S., ``{DIPOL-2: a double image high precision polarimeter},'' in [{\em Ground-based and Airborne Instrumentation for Astronomy V}{\nolinebreak\hspace{0.1em}]},  Ramsay, S.~K., McLean, I.~S., and Takami, H., eds.,  {\bf 9147},  2719 -- 2727, International Society for Optics and Photonics, SPIE (2014).

\bibitem{SpUpNIC}
Crause, L.~A., Gilbank, D., van Gend, C., Worters, H.~L., Sass, C., Kotze, E.~J., Potter, S., Sickafoose, A., Sefako, R., Southworth, J., Macri, L., Thorstensen, J., Galan, C., Skelton, P., Engelbrecht, C., Braker, I., Winkler, H., Pieńkowski, D., S{\"u}rgit, D., Erdem, A., and Burleigh, M., ``{SpUpNIC (Spectrograph Upgrade: Newly Improved Cassegrain): a versatile and efficient low- to medium-resolution, long-slit spectrograph on the South African Astronomical Observatory’s 1.9-m telescope},'' {\em Journal of Astronomical Telescopes, Instruments, and Systems}~{\bf 5}(2),  024007 (2019).

\bibitem{HARPSPol}
{Piskunov}, N., {Snik}, F., {Dolgopolov}, A., {Kochukhov}, O., {Rodenhuis}, M., {Valenti}, J., {Jeffers}, S., {Makaganiuk}, V., {Johns-Krull}, C., {Stempels}, E., and {Keller}, C., ``{HARPSpol {\textemdash} The New Polarimetric Mode for HARPS},'' {\em The Messenger}~{\bf 143},  7--10 (Mar. 2011).

\bibitem{FiberPol_SPIE_2024}
Maharana, S., Chattopadhyay, S., and Bershady, M., ``{FiberPol-6D: spectropolarimetric integral field mode for SAAO 1.9 m telescope using fibers},'' in [{\em Ground-based and Airborne Instrumentation for Astronomy X}{\nolinebreak\hspace{0.1em}]},  Bryant, J.~J., Motohara, K., and Vernet, J. R.~D., eds.,  {\bf 13096},  130969U, International Society for Optics and Photonics, SPIE (2024).

\bibitem{FRD_Sabyasachi_paper}
Chattopadhyay, S., Bershady, M.~A., Wolf, M.~J., and Smith, M.~P., ``{Optimum telescope focal ratios for microlens-to-fiber coupled integral field spectrographs},'' {\em Journal of Astronomical Telescopes, Instruments, and Systems}~{\bf 8}(2),  025001 (2022).

\bibitem{cikota_standards}
Cikota, A., Patat, F., Cikota, S., and Faran, T., ``Linear spectropolarimetry of polarimetric standard stars with vlt/fors2,'' {\em Monthly Notices of the Royal Astronomical Society}~{\bf 464},  4146--4159 (10 2016).

\bibitem{southern_standards_gil-hutton}
Gil-Hutton, R. and Benavidez, P., ``Southern stars that can be used as unpolarized standards,'' {\em Monthly Notices of the Royal Astronomical Society}~{\bf 345},  97--99 (10 2003).

\bibitem{berdyugin_exoplanet}
Berdyugin, A.~V., Berdyugina, S.~V., and Piirola, V., ``{High-precision and high-accuracy polarimetry of exoplanets},'' in [{\em Ground-based and Airborne Instrumentation for Astronomy VII}{\nolinebreak\hspace{0.1em}]},  Evans, C.~J., Simard, L., and Takami, H., eds.,  {\bf 10702},  107024Z, International Society for Optics and Photonics, SPIE (2018).

\end{thebibliography}
\bibliographystyle{spiebib} 

\end{document}